\documentclass[aps,prl,twocolumn,superscriptaddress,showpacs]{revtex4}

\usepackage{graphicx}
\usepackage{dcolumn}
\usepackage{bm}
\usepackage[abs]{overpic}
\usepackage{lineno}

\begin{document}
\newlength{\ogltemp}

\def\etmiss{\big\slash\hspace{-1.6ex}{E_\mathrm{T}}}
\def\htmiss{\big\slash\hspace{-1.6ex}{H_\mathrm{T}}}
\def\ptmiss{\big\slash\hspace{-1.6ex}{P_\mathrm{T}^\mathrm{tr}}}
\def\vmet{\settowidth{\ogltemp}{\big\slash}%
  \hspace{0.5mm}\big\slash\hspace{-1.4\ogltemp}{\vec{E}_\mathrm{T}}}

\def\gev{\,\,\mathrm{GeV}\,} 
\def\tev{\,\,\mathrm{TeV}\,}
\def\mev{\,\,\mathrm{MeV}\,} 
\def\gevc{\,\,\mathrm{GeV}/c\,}
\def\gevm{\,\,\mathrm{GeV}/c^2\,} 
\newcommand{\boldit}[1]{\textbf{\emph{#1}}}

\title{A Search for the Higgs Boson Using Neural Networks in Events with
  Missing Energy and \boldit{b}-quark Jets in $p\overline p$ Collisions
  at $\sqrt{s}=1.96$ TeV}

\affiliation{Institute of Physics, Academia Sinica, Taipei, Taiwan 11529, Republic of China} 
\affiliation{Argonne National Laboratory, Argonne, Illinois 60439} 
\affiliation{University of Athens, 157 71 Athens, Greece} 
\affiliation{Institut de Fisica d'Altes Energies, Universitat Autonoma de Barcelona, E-08193, Bellaterra (Barcelona), Spain} 
\affiliation{Baylor University, Waco, Texas  76798} 
\affiliation{Istituto Nazionale di Fisica Nucleare Bologna, $^{cc}$University of Bologna, I-40127 Bologna, Italy} 
\affiliation{Brandeis University, Waltham, Massachusetts 02254} 
\affiliation{University of California, Davis, Davis, California  95616} 
\affiliation{University of California, Los Angeles, Los Angeles, California  90024} 
\affiliation{University of California, San Diego, La Jolla, California  92093} 
\affiliation{University of California, Santa Barbara, Santa Barbara, California 93106} 
\affiliation{Instituto de Fisica de Cantabria, CSIC-University of Cantabria, 39005 Santander, Spain} 
\affiliation{Carnegie Mellon University, Pittsburgh, PA  15213} 
\affiliation{Enrico Fermi Institute, University of Chicago, Chicago, Illinois 60637}
\affiliation{Comenius University, 842 48 Bratislava, Slovakia; Institute of Experimental Physics, 040 01 Kosice, Slovakia} 
\affiliation{Joint Institute for Nuclear Research, RU-141980 Dubna, Russia} 
\affiliation{Duke University, Durham, North Carolina  27708} 
\affiliation{Fermi National Accelerator Laboratory, Batavia, Illinois 60510} 
\affiliation{University of Florida, Gainesville, Florida  32611} 
\affiliation{Laboratori Nazionali di Frascati, Istituto Nazionale di Fisica Nucleare, I-00044 Frascati, Italy} 
\affiliation{University of Geneva, CH-1211 Geneva 4, Switzerland} 
\affiliation{Glasgow University, Glasgow G12 8QQ, United Kingdom} 
\affiliation{Harvard University, Cambridge, Massachusetts 02138} 
\affiliation{Division of High Energy Physics, Department of Physics, University of Helsinki and Helsinki Institute of Physics, FIN-00014, Helsinki, Finland} 
\affiliation{University of Illinois, Urbana, Illinois 61801} 
\affiliation{The Johns Hopkins University, Baltimore, Maryland 21218} 
\affiliation{Institut f\"{u}r Experimentelle Kernphysik, Karlsruhe Institute of Technology, D-76131 Karlsruhe, Germany} 
\affiliation{Center for High Energy Physics: Kyungpook National University, Daegu 702-701, Korea; Seoul National University, Seoul 151-742, Korea; Sungkyunkwan University, Suwon 440-746, Korea; Korea Institute of Science and Technology Information, Daejeon 305-806, Korea; Chonnam National University, Gwangju 500-757, Korea; Chonbuk National University, Jeonju 561-756, Korea} 
\affiliation{Ernest Orlando Lawrence Berkeley National Laboratory, Berkeley, California 94720} 
\affiliation{University of Liverpool, Liverpool L69 7ZE, United Kingdom} 
\affiliation{University College London, London WC1E 6BT, United Kingdom} 
\affiliation{Centro de Investigaciones Energeticas Medioambientales y Tecnologicas, E-28040 Madrid, Spain} 
\affiliation{Massachusetts Institute of Technology, Cambridge, Massachusetts  02139} 
\affiliation{Institute of Particle Physics: McGill University, Montr\'{e}al, Qu\'{e}bec, Canada H3A~2T8; Simon Fraser University, Burnaby, British Columbia, Canada V5A~1S6; University of Toronto, Toronto, Ontario, Canada M5S~1A7; and TRIUMF, Vancouver, British Columbia, Canada V6T~2A3} 
\affiliation{University of Michigan, Ann Arbor, Michigan 48109} 
\affiliation{Michigan State University, East Lansing, Michigan  48824}
\affiliation{Institution for Theoretical and Experimental Physics, ITEP, Moscow 117259, Russia} 
\affiliation{University of New Mexico, Albuquerque, New Mexico 87131} 
\affiliation{Northwestern University, Evanston, Illinois  60208} 
\affiliation{The Ohio State University, Columbus, Ohio  43210} 
\affiliation{Okayama University, Okayama 700-8530, Japan} 
\affiliation{Osaka City University, Osaka 588, Japan} 
\affiliation{University of Oxford, Oxford OX1 3RH, United Kingdom} 
\affiliation{Istituto Nazionale di Fisica Nucleare, Sezione di Padova-Trento, $^{dd}$University of Padova, I-35131 Padova, Italy} 
\affiliation{LPNHE, Universite Pierre et Marie Curie/IN2P3-CNRS, UMR7585, Paris, F-75252 France} 
\affiliation{University of Pennsylvania, Philadelphia, Pennsylvania 19104}
\affiliation{Istituto Nazionale di Fisica Nucleare Pisa, $^{ee}$University of Pisa, $^{ff}$University of Siena and $^{gg}$Scuola Normale Superiore, I-56127 Pisa, Italy} 
\affiliation{University of Pittsburgh, Pittsburgh, Pennsylvania 15260} 
\affiliation{Purdue University, West Lafayette, Indiana 47907} 
\affiliation{University of Rochester, Rochester, New York 14627} 
\affiliation{The Rockefeller University, New York, New York 10021} 
\affiliation{Istituto Nazionale di Fisica Nucleare, Sezione di Roma 1, $^{hh}$Sapienza Universit\`{a} di Roma, I-00185 Roma, Italy} 

\affiliation{Rutgers University, Piscataway, New Jersey 08855} 
\affiliation{Texas A\&M University, College Station, Texas 77843} 
\affiliation{Istituto Nazionale di Fisica Nucleare Trieste/Udine, I-34100 Trieste, $^{ii}$University of Trieste/Udine, I-33100 Udine, Italy} 
\affiliation{University of Tsukuba, Tsukuba, Ibaraki 305, Japan} 
\affiliation{Tufts University, Medford, Massachusetts 02155} 
\affiliation{Waseda University, Tokyo 169, Japan} 
\affiliation{Wayne State University, Detroit, Michigan  48201} 
\affiliation{University of Wisconsin, Madison, Wisconsin 53706} 
\affiliation{Yale University, New Haven, Connecticut 06520} 
\author{T.~Aaltonen}
\affiliation{Division of High Energy Physics, Department of Physics, University of Helsinki and Helsinki Institute of Physics, FIN-00014, Helsinki, Finland}
\author{J.~Adelman}
\affiliation{Enrico Fermi Institute, University of Chicago, Chicago, Illinois 60637}
\author{B.~\'{A}lvarez~Gonz\'{a}lez$^v$}
\affiliation{Instituto de Fisica de Cantabria, CSIC-University of Cantabria, 39005 Santander, Spain}
\author{S.~Amerio$^{dd}$}
\affiliation{Istituto Nazionale di Fisica Nucleare, Sezione di Padova-Trento, $^{dd}$University of Padova, I-35131 Padova, Italy} 

\author{D.~Amidei}
\affiliation{University of Michigan, Ann Arbor, Michigan 48109}
\author{A.~Anastassov}
\affiliation{Northwestern University, Evanston, Illinois  60208}
\author{A.~Annovi}
\affiliation{Laboratori Nazionali di Frascati, Istituto Nazionale di Fisica Nucleare, I-00044 Frascati, Italy}
\author{J.~Antos}
\affiliation{Comenius University, 842 48 Bratislava, Slovakia; Institute of Experimental Physics, 040 01 Kosice, Slovakia}
\author{G.~Apollinari}
\affiliation{Fermi National Accelerator Laboratory, Batavia, Illinois 60510}
\author{A.~Apresyan}
\affiliation{Purdue University, West Lafayette, Indiana 47907}
% ...

\collaboration{CDF Collaboration\footnote{With visitors from $^a$University of Massachusetts Amherst, Amherst, Massachusetts 01003,
$^b$Universiteit Antwerpen, B-2610 Antwerp, Belgium, 
$^c$University of Bristol, Bristol BS8 1TL, United Kingdom,
$^d$Chinese Academy of Sciences, Beijing 100864, China, 
$^e$Istituto Nazionale di Fisica Nucleare, Sezione di Cagliari, 09042 Monserrato (Cagliari), Italy,
$^f$University of California Irvine, Irvine, CA  92697, 
$^g$University of California Santa Cruz, Santa Cruz, CA  95064, 
$^h$Cornell University, Ithaca, NY  14853, 
$^i$University of Cyprus, Nicosia CY-1678, Cyprus, 
$^j$University College Dublin, Dublin 4, Ireland,
$^k$University of Edinburgh, Edinburgh EH9 3JZ, United Kingdom, 
$^l$University of Fukui, Fukui City, Fukui Prefecture, Japan 910-0017
$^m$Kinki University, Higashi-Osaka City, Japan 577-8502
$^n$Universidad Iberoamericana, Mexico D.F., Mexico,
$^o$University of Iowa, Iowa City, IA  52242,
$^p$Kansas State University, Manhattan, KS 66506
$^q$Queen Mary, University of London, London, E1 4NS, England,
$^r$University of Manchester, Manchester M13 9PL, England,
$^s$Muons, Inc., Batavia, IL 60510, 
$^t$Nagasaki Institute of Applied Science, Nagasaki, Japan, 
$^u$University of Notre Dame, Notre Dame, IN 46556,
$^v$University de Oviedo, E-33007 Oviedo, Spain, 
$^w$Texas Tech University, Lubbock, TX  79609, 
$^x$IFIC(CSIC-Universitat de Valencia), 56071 Valencia, Spain,
$^y$Universidad Tecnica Federico Santa Maria, 110v Valparaiso, Chile,
$^z$University of Virginia, Charlottesville, VA  22906
$^{aa}$Bergische Universit\"at Wuppertal, 42097 Wuppertal, Germany,
$^{bb}$Yarmouk University, Irbid 211-63, Jordan
$^{jj}$On leave from J.~Stefan Institute, Ljubljana, Slovenia, 
}}
\noaffiliation


\begin{abstract}

  We report on a search for the standard model Higgs boson produced in
  association with a $W$ or $Z$ boson in $p\overline{p}$ collisions at
  $\sqrt{s} = 1.96$ TeV recorded by the CDF II experiment at the
  Tevatron in a data sample corresponding to an integrated luminosity of
  2.1\,fb$^{-1}$. We consider events which have no identified charged
  leptons, an imbalance in transverse momentum, and two or three jets
  where at least one jet is consistent with originating from the decay of a $b$
  hadron. We find good agreement between data and background predictions. We place
  95\% confidence level upper limits on the production cross section for
  several Higgs boson masses ranging from 110$\gevm$ to 150$\gevm$. For
  a mass of 115$\gevm$ the observed (expected) limit is 6.9 (5.6) times
  the standard model prediction.
\end{abstract}
\maketitle

The Higgs boson is the last particle of the standard model (SM) of
particle physics that remains to be discovered. The existence of
the Higgs boson is expected to be the direct physical manifestation
of the mechanism that provides mass to fundamental
particles~\cite{higgs}. Expectations from electroweak data collected
at the Tevatron, LEP, and SLD indirectly constrain the Higgs boson
mass to $m_H< 157$$\gevm$ at 95\% confidence level
(C.L.)~\cite{precisemeas}. Direct searches at LEP have excluded
$m_H< 114.4$$\gevm$ at 95\% C.L.~\cite{lepdirect}. Upper limits on
the production cross section from searches in the region
$110<m_H<135$ $\gevm$ remain well above the standard model
prediction~\cite{precisemeas}, and greatly benefit from improvements of the
experimental sensitivity. In this mass region, $b\overline{b}$ is the main decay
mode. The $b$ quarks fragment into jets of hadrons, and the signal
can be reconstructed as a resonance in the invariant mass
distribution of the two jets. Large multi-jet backgrounds can be reduced by searching for a
Higgs boson ($H$) with an associated vector boson $V$ ($V = W$, $Z$).

This Letter presents a search for the standard model \textit{VH}
associated production in events with $b$-quark jets and large missing
transverse energy with data corresponding to an integrated luminosity of
2.1\,fb$^{-1}$. This analysis significantly increases the acceptance for
signal with respect to previous Tevatron
searches~\cite{cdfmetbb1fb,metbbdzero} and introduces advanced analysis
methods. We consider \textit{ZH} production, where $Z\rightarrow \nu
\bar{\nu}$ and the neutrinos ($\nu$) escape detection, or $Z\rightarrow
\ell \ell$ when both charged leptons ($\ell$) are undetected or are
identified as jets. For \textit{WH} production we are sensitive to
events where $W\rightarrow e\nu$ or $W\rightarrow\tau\nu$ when the
charged lepton is identified as a jet, and $W\rightarrow\ell\nu$ when
$\ell$ is undetected. The $WH$ events accepted by this analysis contain
50\% $W\rightarrow\tau\nu$, 30\% $W\rightarrow\mu\nu$ and 20\%
$W\rightarrow e\nu$. Critical challenges for this analysis are to
achieve a high signal-to-background (S/B) ratio and to model the multi-jet
background production accurately. We employ artificial neural networks
(ANNs)~\cite{neuralnets} to improve the event selection and signal
discrimination and implement a novel data-driven determination of the
multi-jet background.

CDF II is a multipurpose detector that is described in detail
elsewhere~\cite{CDFII,coord}. Jets are reconstructed from energy
depositions in the calorimeter towers using a jet clustering cone
algorithm~\cite{cdfjets} with a cone size of radius $\Delta R
=\sqrt{(\Delta\phi)^2 + (\Delta\eta)^2} = 0.4$. In addition to standard
jet energy corrections~\cite{cdfjets}, we further correct using momentum
measurements provided by the tracker, using a method similar to that
described in~\cite{H1}. The more precise measurement of the jet energies
improves the candidate Higgs boson mass resolution by $\approx$\,10\%
and increases the signal acceptance by $\approx$\,10\%. Both the
magnitude and direction of $\etmiss$ are recomputed after the jet
energies are corrected.

The events used in this search are selected by a three-level trigger
system that selects events with $\etmiss$ and two jet clusters. After offline
reconstruction, the event selection requires $\etmiss>50$\,GeV, and the
transverse energies $E_{\mathrm{T}}^{J_1}$ and $E_{\mathrm{T}}^{J_2}$ of
the two jets with the highest transverse energy, $J_1$ and $J_2$
(``leading jets''), satisfy the conditions
$E_{\mathrm{T}}^{J_1}>35$\,GeV and $E_{\mathrm{T}}^{J_2}>25$\,GeV. We
consider an event to have three jets if the $E_{\mathrm{T}}$ of the
third leading jet, $J_3$, is greater than $15$\,GeV. Events with four or
more jets with $E_{\mathrm{T}} > 15$\,GeV and $|\eta| < 2.4 $ are
rejected. Events passing these criteria are denoted as the
``pre-tagged'' sample.  After these selections, the expected
S/B ratio is around $1/20\,000$.

We veto events with at least one $p_{\mathrm{T}}>10\gevc$ isolated
electron or muon~\cite{cdfmetbb2.1fb}, deliberately using fairly loose
identification criteria. These selections ensure that the sample used in
this analysis is statistically independent from the one utilized in the
search for $WH$ in a final state containing an identified charged
lepton~\cite{cdfwh}.

Large backgrounds originating
from light-flavor jet production can be reduced by identifying $b$ jets
in the candidate events. Due to their relatively long lifetime, $b$ and
$c$ hadrons can travel a few millimeters from the primary vertex before
decaying into lighter hadrons. Jets originating from a $b$ quark can be
identified (``tagged'') by the {\sc secvtx} algorithm~\cite{secvtx},
which reconstructs vertices that are significantly displaced from the
$p\overline{p}$ interaction point, and the {\sc jetprob}
algorithm~\cite{jetprob}, which classifies jets using the probability
that tracks within the jet are consistent with originating from the
primary vertex. To enhance the expected signal significance we subdivide
the sample into three independent categories: events with two jets
tagged by {\sc secvtx} (SV+SV), events with one jet tagged by {\sc
  secvtx} and another by {\sc jetprob} (SV+JP) and events with only one
jet tagged by {\sc secvtx} (SV).

The selected sample is dominated by background from the production of
multi-jet (MJ), top quark (pair and electroweak production), $W$ or $Z$
plus jets, and $WW$, $WZ$ or $ZZ$ events. Significant $\etmiss$ in
multi-jet events appears when $b$ quarks decay semileptonically or when
jet energies are mismeasured. In both cases $\vec \etmiss$ is often
aligned with $\vec E_{\mathrm{T}}^{J_2}$. Therefore, events with $\Delta
\phi (\vec\etmiss,\vec E_{\mathrm{T}}^{J_2})<0.4$, no identified
leptons, and $50<\etmiss<70$\,GeV are used to measure the rates with
which the heavy-flavor jets (h.f., originating from a $b$ or $c$ quark)
from multi-jet production and light flavor jets mistakenly identified as
$b$ jets (``mistags'') are tagged.  The tagging rate ({\sc TR}) is
parametrized as a function of $H_{\mathrm{T}}$~\cite{coord} of the event
and $E_{\mathrm{T}}$, $\eta$, and $\zeta$ of the jet. The observable
$\zeta$ is defined as
$\zeta=c\sum_ip^i_{\mathrm{T,track}}/E_{\mathrm{T}}$ where
$p^i_{\mathrm{T,track}}$ includes tracks within a jet with a significant
impact parameter and $0.5<p^{i}_{\mathrm{T,track}}<200\gevc$. Jets
originating from $b$ quarks are expected to have a large $\zeta$. The
multi-jet background in the single (double) tagged sample is determined
by the probability to tag one (two) jet(s) from the pre-tagged
(single-tagged) sample~\cite{cdfmetbb2.1fb, cdfmetbb_singletop}, after
subtracting all Monte Carlo (MC) simulated contributions. The validity
of the tagging rate modeling is verified in various control regions,
which are defined below. The remaining backgrounds are estimated using
{\sc pythia}~\cite{pythia} simulations, and single top production is
simulated with {\sc madevent}~\cite{madevent}. The signal MC samples are
generated with {\sc pythia}. The normalizations of the MC samples are
described in~\cite{cdfmetbb2.1fb}.

 \begin{figure}
   \includegraphics[width=6.8cm,angle=90]{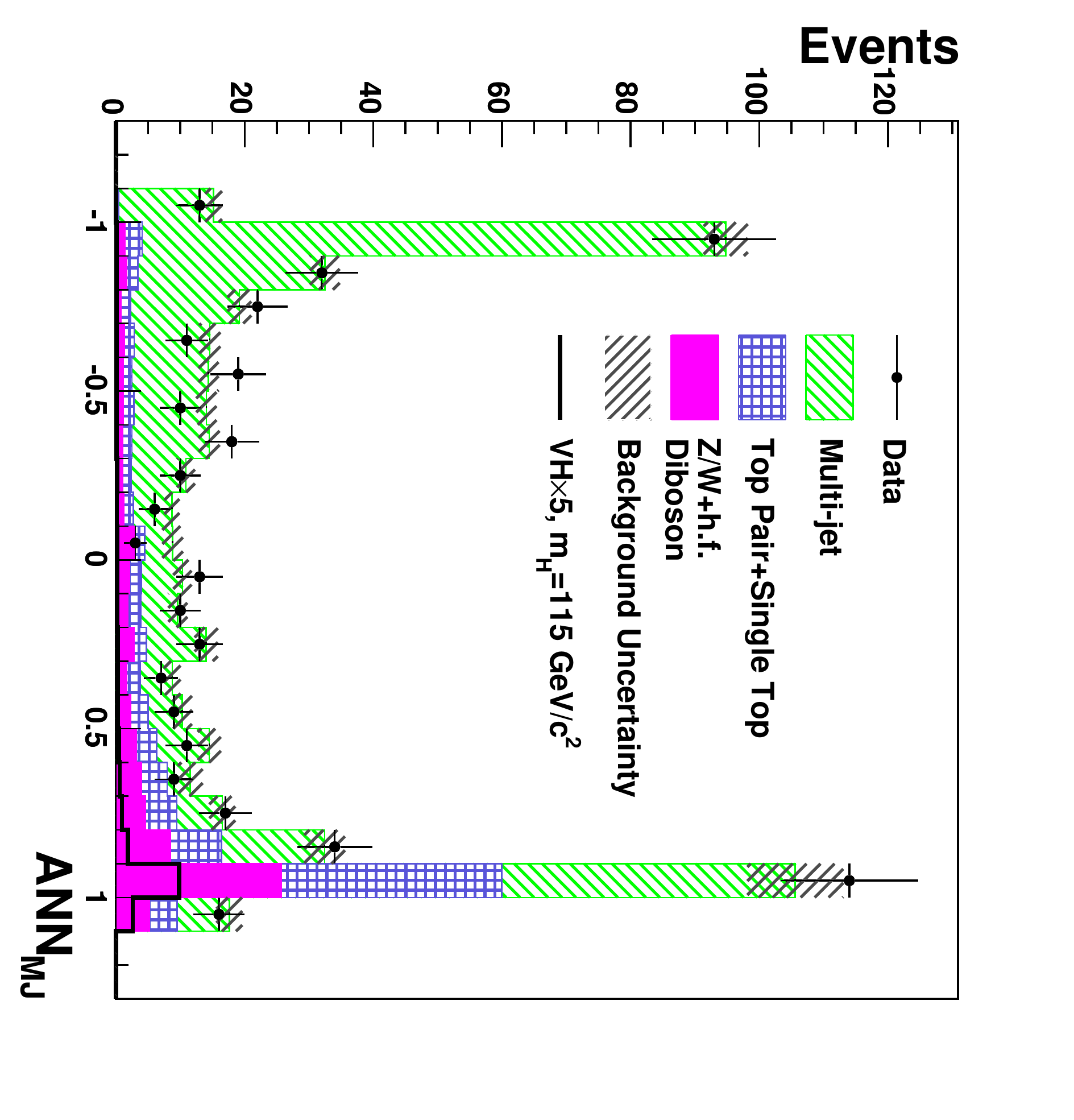}
   \caption{The distribution of ANN$_\mathrm{MJ}$ for double-tagged events.}
   \label{fig:QCDNN}
 \end{figure}

 We start the selection of the signal region by requiring no identified
 charged leptons, $\etmiss>50$\,GeV, $\Delta\phi(\vec{\etmiss},\vec
 E_{\mathrm{T}}^{J_1})\geq1.5$, and $\Delta\phi(\vec \etmiss ,\vec
 E_{\mathrm{T}}^{J_{2,3}})\geq0.4$.  These selection criteria remove
 $\approx$\,10\% of the signal in the pretag sample while reducing the
 backgrounds by approximately an order of magnitude. We employ an {\sc
   ANN}, denoted as ANN$_\mathrm{MJ}$, using kinematic variables to
 separate signal from multi-jet background. To discriminate against
 events with $\etmiss$ due to mismeasurements in the calorimeter, we use
 the momentum imbalance in the tracker, $\ptmiss$~\cite{coord}. The
 magnitude of $\etmiss$, $\ptmiss$, the angle between them, the
 azimuthal angles between $\etmiss$, $\ptmiss$ and the jet directions,
 and several other less discriminating variables are used as inputs to
 ANN$_\mathrm{MJ}$~\cite{cdfmetbb2.1fb}. The distribution of
 ANN$_\mathrm{MJ}$, shown in Fig.\,\ref{fig:QCDNN}, peaks at $+1$ for
 the signal and at $-1$ for the backgrounds that are due to mismeasured
 jets. Selecting events with ANN$_\mathrm{MJ}\geq 0$ rejects over 50\%
 of the total background and retains 95\% of the signal, yielding an S/B
 ratio of~$\approx$~1/250, which is similar to the one obtained in the
 WH search~\cite{cdfwh}. This region is defined as the signal region and
 is analyzed for the presence of the Higgs boson signal.

In order to avoid potential bias in the search, we test our
understanding of the SM background in several control regions where the
amount of signal events is negligible. The
control region called EWK, sensitive to electroweak processes and top
production, contains events with at least one lepton and
$\Delta\phi(\vec{\etmiss},\vec E_{\mathrm{T}}^{J_2})\geq 0.4$. We also
define several control regions dominated by multi-jet processes where we
have no identified leptons. The region denoted as MJ1 contains events
with $\Delta\phi(\vec{\etmiss},\vec E_{\mathrm{T}}^{J_2})< 0.4$ and
$\etmiss\geq 70$\,GeV. The region denoted as MJ2 contains events with
$\Delta\phi(\vec{\etmiss},\vec E_{\mathrm{T}}^{J_1})\geq 1.5$,
$\Delta\phi(\vec{\etmiss},\vec E_{\mathrm{T}}^{J_{2,3}})\geq 0.4$, and
ANN$_\mathrm{MJ}<-0.5$. The predictions of the multi-jet background are
tested in MJ1 and MJ2. The distributions of kinematic variables, such as
the invariant mass of the two leading jets $m({J_1,J_2})$, have been
found to be in agreement with observations in the control
regions~\cite{cdfmetbb2.1fb}.

\begin{table}
  \caption{Comparison of the total number of expected and observed
    events in the signal region for
    different $b$-tagging categories. The uncertainties contain both
    MC statistical error and systematic uncertainties.}
  \label{tab:table1}
    \begin{tabular}{lccc}\hline\hline
      Process              & SV+SV or SV+JP & SV        \\
      \hline
      Multi-jet             & $ 120.1\pm 21.3 $ & $ 941.2 \pm 86.0  $  \\
      Single Top           & $ 15.7 \pm 3.0  $ & $ 43.2  \pm 7.9   $  \\
      Top Pair             & $ 54.5 \pm 7.9  $ & $ 124.5 \pm 17.0  $  \\
      Di-boson             & $ 9.2  \pm 1.8  $ & $ 35.6  \pm 6.8   $  \\
      W + h.f.             & $ 32.0 \pm 14.7 $ & $ 296.9 \pm 129.5 $  \\
      Z + h.f.             & $ 22.1 \pm 11.5 $ & $ 107.0 \pm 45.8  $  \\
      \hline
      Total Prediction     & $ 254  \pm 39  $ & $ 1548.4 \pm 168.1 $  \\ \hline
      Observed             & $ 253$           & $ 1443             $  \\ \hline\hline
     \multicolumn{3}{l}{Expected Signal for $m_H$=115$\gevm$} \\\hline
      $ZH\rightarrow\nu\nu bb$     & $ 1.8 \pm 0.2 $ & $ 2.1 \pm 0.2 $ \\
      $WH\rightarrow(\ell)\nu bb$  & $ 1.6 \pm 0.2 $ & $ 1.8 \pm 0.2 $ \\
      $ZH\rightarrow(\ell\ell) bb$ & $ 0.07\pm 0.01$ & $ 0.09\pm 0.01$ \\\hline\hline
    \end{tabular}
\end{table}

\begin{figure}
  \includegraphics[width=6.8cm,angle=90]{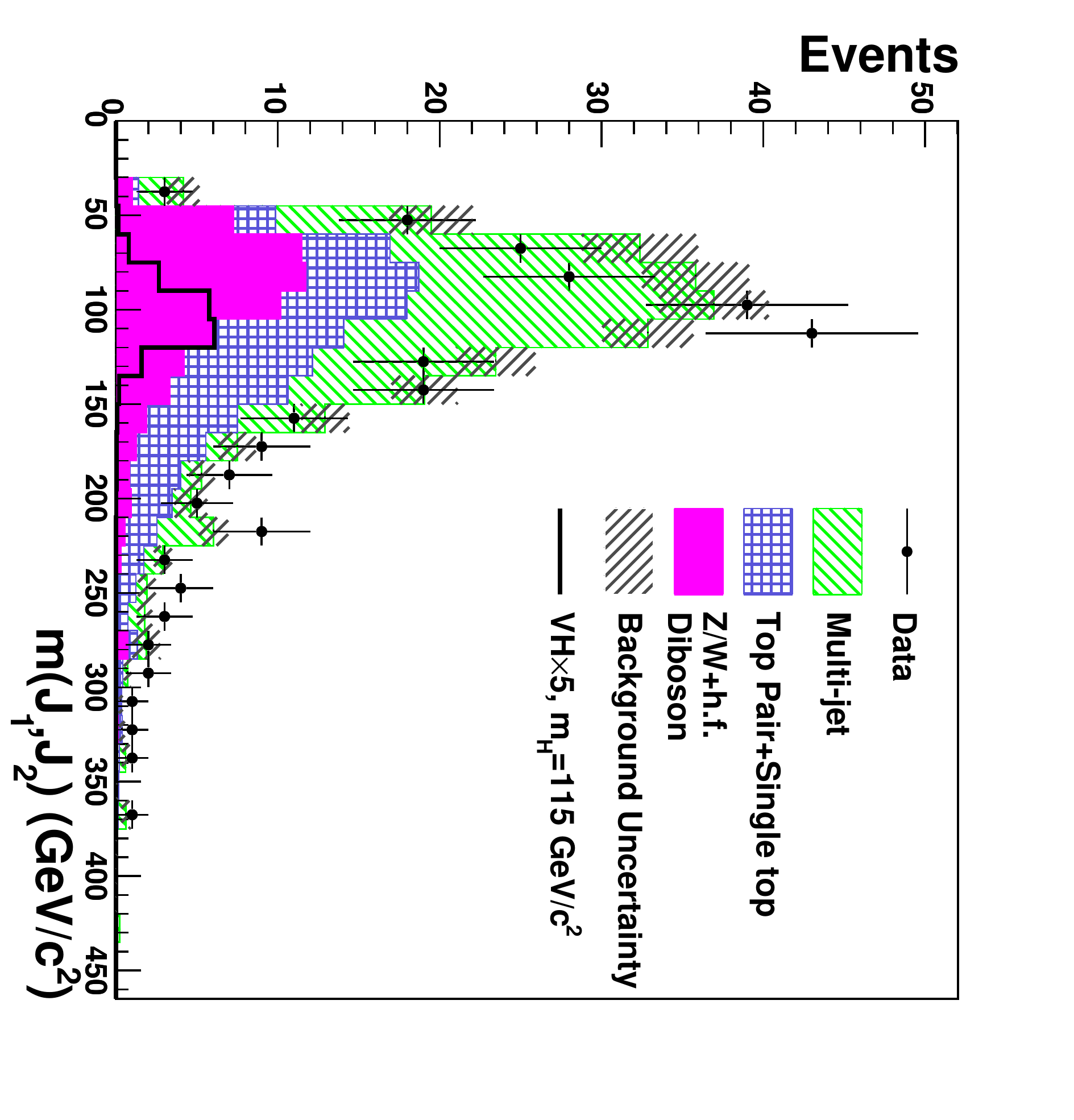}
 \caption{Dijet invariant mass distribution for double-tagged events in
    signal region.}
  \label{fig:Mjj}
\end{figure}
 
To achieve a greater separation between signal and background we deploy
a second ANN, denoted as ANN$_\mathrm{sig}$, for discriminating the
remaining backgrounds from the expected signal. Six input variables are
used in ANN$_\mathrm{sig}$: the invariant mass of the two leading jets
$m({J_1,J_2})$~(Fig.\,\ref{fig:Mjj}), the invariant mass of the $\etmiss$ and all jets,
$H_{\mathrm{T}}-\etmiss$, $\htmiss-\etmiss$, {\sc
  trackmet}~\cite{metbbnn} and the maximum $\Delta R(\vec
E_{\mathrm{T}}^{J_i},\vec E_{\mathrm{T}}^{J_k})$.  The variable {\sc
  trackmet} is the output of a ANN developed using tracking information
to enhance the separation of events with real $\etmiss$. The most
discriminating variable of the ANN$_\mathrm{sig}$ is $m({J_1,J_2})$.

\begin{figure}
 \begin{overpic}[width=6.3cm]{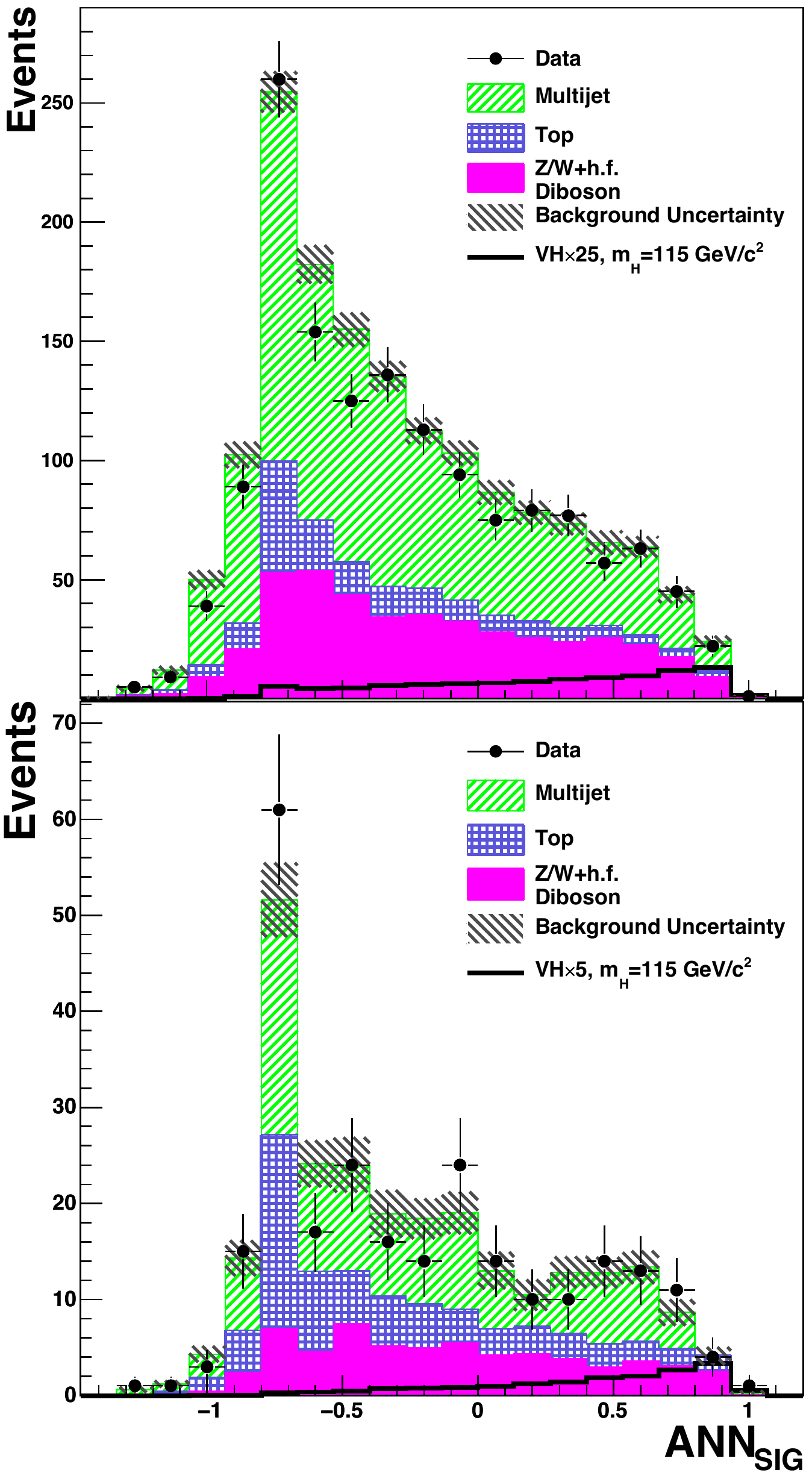}\put(22,315){\textbf{(a)}}\put(22,161){\textbf{(b)}} \end{overpic}
 \caption{The distribution of ANN$_\mathrm{sig}$ for (a) single- and (b) double-tagged events.}
  \label{fig:SIGNN}
\end{figure}
 
The distribution of ANN$_\mathrm{sig}$ is shown in Fig.~\ref{fig:SIGNN}
for single- and double-tagged events. The number of signal and background events
after the final selection are shown in Table \ref{tab:table1}. Since no
significant excess is observed, we compute 95\% C.L. upper limits for
the Higgs boson production cross section times the branching
fraction $B(H \to b\bar{b})$. For $m_H$=115$\gevm$ we expect a total of 4.0 (3.5) signal
events with one (two) $b$-tagged jets~\cite{Higgs}.

We analyze the binned ANN$_\mathrm{sig}$ discriminant distribution to
test for a \textit{WH} or \textit{ZH} signal in the presence of SM
backgrounds. The systematic uncertainties included in the calculation
are classified as correlated (uncorrelated) depending on if they do (do
not) affect both signal and the background
processes~\cite{cdfmetbb2.1fb,metbbnn}. The uncorrelated systematic
uncertainties are the multi-jet normalization (between 5.5\% and 20.6\%)
and MC statistical fluctuations. Additionally we assign the following
uncertainties due to cross sections: 15.9\% and 15.2\% to single top in
$s$- and $t$-channels, 8.6\% to top pair, 11.5\% to diboson and 40\% to
$W$ + h.f. and $Z$ + h.f. The shapes obtained by varying the {\sc TR}
probabilities by $\pm 1\sigma$ are applied as systematic uncertainties
to each bin of ANN$_\mathrm{sig}$. The correlated systematic
uncertainties are the following: luminosity measurement (6.0\%),
$b$-tagging efficiency in MC (between 4.3\% and 12.4\%), trigger
efficiency ($<$3\%), lepton veto efficiency (2\%), parton distribution
function uncertainty (2\%) and 3.8-13.0\% for jet energy scale
(JES)\,\cite{cdfjets}. We also assign systematic uncertainties on the
shape of ANN$_\mathrm{sig}$ due to JES and trigger efficiency
uncertainties. Initial- and final-state-radiation systematic
uncertainties (between 1\% and 5\%) are applied to the signal
predictions.

Including all the uncertainties, the expected and observed upper
limits at the 95\% C.L. on \textit{VH} production cross section
times branching fraction $Br(H \to b\bar{b})$ are shown in Table
\ref{tab:table2}. Expected limits are obtained by generating
pseudo-experiments from the expected SM ANN$_\mathrm{sig}$ shapes
to calculate the median \textit{ZH} and \textit{WH} contribution
which could be excluded at the 95\% C.L. in the zero signal
hypothesis.  The limits are computed using the Bayesian likelihood
method~\cite{limits} with a flat prior probability for the signal
cross section and Gaussian priors for the uncertainties on
acceptance and backgrounds.  We combine the search channels SV+SV,
SV+JP, and SV by taking the product of their likelihoods and
simultaneously varying the correlated uncertainties. The observed limits agree with the
expected ones.

\begin{table}
  \caption{Expected and observed 95\% C.L. upper limits, with ratios to SM cross section.}
  \label{tab:table2}
  \ruledtabular
  \begin{tabular}{lcccc}
    $m_H$ & Expected   & Observed & Ratio    & Ratio     \\
    ($\mathrm{GeV}/c^2$)    &  (pb)      & (pb)     & Expected & Observed  \\

    \hline\hline
    110   & $1.3$    & $1.5$  & $4.9^{+2.1}_{-1.4}$    & 5.8 \\
    115   & $1.2$    & $1.5$  & $5.6^{+2.4}_{-1.6}$    & 6.9 \\
    120   & $1.2$    & $1.5$  & $7.2^{+2.9}_{-2.1}$    & 8.9 \\
    130   & $1.0$    & $1.4$  & $10.3^{+4.3}_{-2.9}$   & 14.4 \\
    140   & $0.9$    & $1.0$  & $18.6^{+7.8}_{-5.4}$   & 21.0 \\
    150   & $0.8$    & $1.0$  & $43.3^{+19.0}_{-12.4}$ & 49.8\\
  \end{tabular}
\end{table}

In summary, we have performed a direct search for the SM Higgs boson
decaying into $b$-jet pairs using data with integrated luminosity of
$2.1$\,fb$^{-1}$ accumulated in Run II by the CDF II detector. The combination of all improvements
described above increases the sensitivity of this search by a factor of
two with respect to~\cite{cdfmetbb1fb}, and by 30\% with respect
to~\cite{metbbdzero} once the difference in luminosity is accounted for.

We thank the Fermilab staff and the technical staffs of the
participating institutions for their vital contributions. This work was
supported by the U.S. Department of Energy and National Science
Foundation; the Italian Istituto Nazionale di Fisica Nucleare; the
Ministry of Education, Culture, Sports, Science and Technology of Japan;
the Natural Sciences and Engineering Research Council of Canada; the
National Science Council of the Republic of China; the Swiss National
Science Foundation; the A.P. Sloan Foundation; the Bundesministerium
f\"ur Bildung und Forschung, Germany; the World Class University
Program, the National Research Foundation of Korea; the Science and
Technology Facilities Council and the Royal Society, UK; the Institut
National de Physique Nucleaire et Physique des Particules/CNRS; the
Russian Foundation for Basic Research; the Ministerio de Ciencia e
Innovaci\'{o}n, and Programa Consolider-Ingenio 2010, Spain; the Slovak
R\&D Agency; and the Academy of Finland.

\bibliography{apssamp}

\begin{thebibliography}{99}
\bibitem{coord} CDF uses a cylindrical coordinate system with the $z$
  axis along the proton beam axis. Pseudorapidity is
  $\eta=-\ln(\tan(\frac{\theta}{2}))$, where $\theta$ is the polar
  angle, and $\phi$ is the azimuthal angle relative to the proton beam
  direction, while \mbox{$p_{\mathrm{T}}=p \sin\theta$},
  \mbox{$E_{\mathrm{T}}=E \sin\theta$}. The $\etmiss$ is defined as the
  magnitude of $\vmet = -\sum_{i} E_{\mathrm{T}}^i\hat{n}_i$, where
  $\hat{n}_i$ is a unit vector perpendicular to the beam axis and
  pointing at the $i^{th}$ calorimeter tower, and $E_{\mathrm{T}}^i$ is
  the transverse energy therein. The scalar sum of transverse energies
  of the leading jets is denoted as $H_{\mathrm{T}}$, and the vector sum
  of jet $E_{\mathrm{T}}$'s is denoted as $\htmiss$. The $\vec\ptmiss$ is defined as negative
  vector sum of track $p_{\mathrm{T}}$'s.
\end{thebibliography}

\end{document}